\RequirePackage{fix-cm}
\documentclass[twocolumn]{svjour3}          
\smartqed  
\usepackage{graphicx}
\begin{document}

\title{Oscillator strengths of the intersubband electronic transitions in the multi-layered nano-antidots with hydrogenic impurity
}

\author{Y. Naimi \and  A. R. Jafari}


\institute{Y. Naimi \and
           A. R. Jafari \at
              Department of Physics, Lamerd Branch, Islamic Azad University, Lamerd, Iran.\\
              Tel.: +98-782-5221100\\
              Fax: +98-782-5221232\\
              \email{y.naimi@iaulamerd.ac.ir}           
           }

\date{Received: date / Accepted: date}

\maketitle

\begin{abstract}
In this study, we have obtained the exact solutions of the Schr\"{o}dinger equation for a multi-layered quantum antidot (MLQAD) within the effective mass approximation and dielectric continuum model for the spherical symmetry. The MLQAD is nano-structured semiconductor system that consists of a spherical core (e.g. $Ga_{1-x}Al_{x}As $) and a coated spherical shell (e.g. $Ga_{1-y}Al_{y}As $) as the whole anti-dot is embedded inside a bulk material (e.g. $GaAs $). The dependence of the electron energy spectrum and its radial probability density on nano-system radius are studied. The numeric calculations and analysis of
oscillator strength of intersubband quantum transition from the ground state into two first allowed excited states at the varying radius, for both the finite and infinite confining potential (CP) as well as constant shell thickness, are performed. It is shown that, in particular, the binding energy and the oscillator strength of the hydrogenic impurity of a MLQAD behave differently from that of a single-layered quantum antidot (SLQAD). For a MLQAD with finite core and shell CPs, the state energies and the oscillator strengths of the impurity are found to be dependent on the shell thickness. At the large core radius and very small shell thickness, our results are closer to respective values for a SLQAD that previously reported.
\end{abstract}

\section{Introduction}
\label{intro}
In the recent years, low-dimensional semiconductors have attracted much attention, due to a wide range of their applications in electronic and optoelectronic devices. The nano-structured semiconductor systems such as quantum dots (QDs), quantum antidots (QADs), wells, well wires and antiwells have been central point of both the theoretical and the experimental researches. One of the most studied classes of heterostructures is the class of QDs. In QDs, energy levels in valance and conduction bands
have discrete distribution and the band gap is larger than compared to bulk state because
of quantum confinement effects. Quantum confinement effects occur when the size of
the structure is comparable to the Bohr radius. In addition to QDs, in the past few years, a great attention has been devoted to the physics of the semiconductor QADs \cite{boga,aqui}.

A QAD is a portion of the semiconductor material that its charge carriers are confined in all three spatial dimensions. So, QADs have  electronic characteristics intermediate between those of bulk semiconductors and those of discrete molecules.  The electronic properties of QADs are closely related to their size and shape. The potential application of QADs in the development of semiconductor optoelectronic devices has led to much study of their electronic characteristics \cite{khor,plan,madhav}.

One of the most studied problems during the last decade is the investigation of properties of semiconductor when it is doped with shallow donor impurities \cite{holo,davat}. The thermal, optical and electrical properties of a doped semiconductor are dependent on impurity states. Consequently, the calculation of impurity states in semiconductors has received considerable attention \cite{kohn,hung}.
 The first systematic study of hydrogenic impurity states in a GaAs quantum well has been done by Bastard \cite{bas1,bas2}.

Many studies of the single-layered QD with a central donor impurity for different confining potential values, have been reported in
\cite{safak,ozmen,sadeghi,osorio,siva}. Varshni in \cite{varshini} calculated the energies for $1s$, $2p$ and $3d$ states of a QD by using of the variational method. The Varshni's results are well consistent with exact solutions that have reported in \cite{poras,Chuu,tkac1,tkac2,zu}.

For multi-layered quantum dot (MLQD) case, consists of a spherical core (e.g. $GaAs$) and a coated spherical shell
(e.g. $Ga_{1-x}Al_{x} As$) as the whole dot is embedded inside a bulk material (e.g. $Ga_{1-y}Al_{y} As$), Cheng-Ying and Der Sun Chuu \cite{cheng} have solved the Schr\"{o}dinger equation and have calculated the exact ground sate energy of a hydrogenic impurity located at the center of MLQD in the framework of the effective mass approximation. Their studies shown that as the dot radius approaches infinity, the state energies of an impurity located at the center of a multi-layered or a single-layered QD approach $-1/n^{2}Ry$, where $n$ is the principle quantum number.

In this paper, we investigate the properties of the multi-layered quantum antidots (MLQAD). The MLQAD consist of a spherical core (e.g. $Ga_{1-x}Al_{x}As $) and a coated spherical shell (e.g. $Ga_{1-y}Al_{y}As $) as the whole antidot is embedded inside a bulk material (e.g. $GaAs $) with a hydrogenic impurity in the center. We obtain the energies and probability density functions for both the ground and first excited state of the MLQAD with different core and shell CPs. Then, by using of  energies and wave functions, the oscillator strength of intersubband transition involving the ground state to the first allowed excited state are computed. Our results show that for small core radii the oscillator strengths are dependent to shell thickness and core CP values but when the core and shell radius are several times of Bohr radius, the oscillator strengths are independent of the shell thickness and core CP.

For many semiconductor quantum heterostructures, such as $GaAl$/$Ga_{1-x}Al_{x}As$, the polarization and image
charge effects can be significant in the multi-layered system if there is a large dielectric
discontinuity between the dot and the surrounding medium. However, this is not the case for
the $GaAs/Ga1_{x}Al_{x}As$ quantum system \cite{ada}, therefore these effects can be ignored safely in our calculation. Furthermore, for the sake of generality, the difference between the electronic effective mass in the antidot core, the shell material and the bulk material has been ignored (i.e., $\mu_1\simeq\mu_2 \simeq\mu_3 =\mu$). The effective atomic units are used throughout the paper so that all energies are measured in units of the effective Rydberg, $\rm{Ry}=\mu e^4/2\epsilon^2\hbar^2$, and distances are in units of the effective Bohr radius, $a_0 =\epsilon\hbar^2/\mu e^2$, where
$\mu$ is the electronic effective mass, and $\epsilon$ is the dielectric constant of the semiconductor medium. For instance, in the particular
 case of $GaAs$-based semiconductors, $\mu=0.067 m_e$, and $\epsilon=13.18$ . Thus, for a GaAs host, the effective Rydberg is numerically
$\rm{Ry}=5.2$  meV, and the effective Bohr radius $a_0=104$ {\AA}.

The rest of this paper is organized as follows. In section $2$, the theory of electron energy states spectrum
and wave functions is developed for the MLQAD with hydrogenic donor impurity at the center. In
theoretical calculations, effective mass approximation and spherically symmetric are employed. In this sense, the exact solution of the Schr\"{o}dinger equation is expressed in terms of the Whittaker functions. In section $3$ the energy spectrum, radial probability distribution and oscillator strength of the intersubband transitions are obtained for different cases. Finally, our conclusions are presented in the last section.

\section{Theory and formulation}
\label{sec:1}
In the effective mass approximation, the Hamiltonian for an on-center hydrogenic impurity in MLQADs is given by
\begin{equation}\label{1} \hat{H} =-\frac{\hbar^{2}}{2\mu}\nabla^2 + U(r) \end{equation}
where $\mu$ is the electronic effective mass in the semiconductor medium and $U(r)$ represents the total potential (coulomb and CP) as following
 \begin{equation} \label{2} U(r) =  \left\{
 \begin{array}{ll}
   V_{1}-\frac{e^{2}}{\epsilon r}, & \hspace{1.5cm}r<a \\
   V_{2}-\frac{e^{2}}{\epsilon r}, &  \hspace{1.5cm} a\leq r \leq b\\
  V_{3}-\frac{e^{2}}{\epsilon r}, & \hspace{1.5cm} r>b
 \end{array}\right.
 \end{equation}
where $a$ is the core radius and $b$ is the total antidot (core plus shell) radius, therefore $b-a$ is the thickness of the
shell. $V_{1}$ and $V_{2}$ are the positive CP inside the core and shell media respectively. The relation between the value of $V_{1}$ and $V_{2}$ is $V_{1}\geq V_{2}$, this condition ensures that the electron reminds bind in several lowest state energies at least. Furthermore, the value of $V_{3}$ is always zero (i.e., the bulk material have no CP) but writing $V_{3}$ helps us to be able for writing the next equations in the compact form as you will see.

In the three dimensions spherical coordinate, the Schr\"{o}dinger eigenvalue equation is $\hat{H} \psi (r,\theta,\phi)=E\psi (r,\theta,\phi)$, where the eigenfunctions are of
the separable form $\psi(r,\theta,\phi)= R(r)Y(\theta,\phi)$. For the $r$-dependent (spherically symmetric) potentials, the angular dependence
of the wave function $Y(\theta, \phi)$, is given by the spherical harmonics \cite{zet}. In particular, the angular part is unaffected by the core and total radius, thus we ignore this part. The radial part is affected by the core and total radius and the antidot CPs, so in the following it is considered.
The radial part of the Schr\"{o}dinger equation in the spherical coordinate for three regions are given by the following compact relation
\begin{equation}\label{3}
\frac{\hbar^{2}}{2\mu}(\frac{d^{2}}{dr^{2}}+\frac{2}{r}\frac{d}{dr}-\frac{l(l+1)}{r^{2}})R^{(i)}(r)+(E-V_{i}+\frac{e^{2}}{\epsilon r})R^{(i)}(r)=0
\end{equation}
where $i=1$, $2$ and $3$ are corresponding to $r<a$, $a<r<b$ and $r>b$ respectively. Using the convenient parameters
\begin{equation}\label{6}\alpha^{2}_{i}=\frac{-8\mu (E-V_{i})}{\hbar^{2}},\ \ \ \ \xi_{i}=\alpha_{i}r,\ \ \ \lambda_{i}=\frac{2\mu e^{2}}{\epsilon\hbar^{2}\alpha_{i}^2}.\end{equation}
and further writing $W(\xi_{i})=\xi_{i} R^{(i)}(r)$, then the above Schr\"{o}dinger equation conveniently converts to
the following whittaker equation
\begin{equation}\label{9}
\frac{d^{2}W(\xi_{i})}{d\xi_i^{2}}+\left(\frac{-1}{4}+\frac{\lambda_{i}}{\xi_{i}}+\frac{\frac{1}{4}-(l+\frac{1}{2})^{2}}{\xi^{2}_{i}}\right)
W(\xi_{i})=0,
\end{equation}
The general solution is the whittaker function \cite{abra}
\begin{equation}\label{12} W(\xi_{i})=C_{1i}M_{\lambda_{i},l+\frac{1}{2}}(\xi_{i})+
C_{2i}W_{\lambda_{i}, l+\frac{1}{2}}(\xi_{i}),\end{equation}
where $C_{1i}$ and $C_{2i}$ are constants to be specified by continuity, asymptotic and normalization conditions.
The asymptotic behavior dictates that the wave function must be finite and zero in origin neighborhood and far away from origin, respectively. These conditions lead to $C_{21}=C_{13}=0$ (since $W_{\lambda_{i}, l+\frac{1}{2}}(\xi_{i})$ and $M_{\lambda_{i},l+\frac{1}{2}}(\xi_{i}$) diverge at the origin and infinity, respectively). Now we can write the wave functions in three regions as
 \begin{eqnarray}
 &&\hspace{-0.2cm}R^{(1)}=C_{11}\frac{M_{\lambda_{1},l+\frac{1}{2}}(\xi_{1})}{\xi_{1}}\ \ \ \hspace{1cm}r<a \\
 &&\hspace{-0.2cm}R^{(2)}=\frac{C_{12}M_{\lambda_{2},l+\frac{1}{2}}(\xi_{2})+ C_{22}W_{\lambda_{2},l+\frac{1}{2}}(\xi_{2})}{\xi_{2}}
 \ \ \ \ a\leq r \leq b \nonumber \\
 && \\&& \hspace{-0.2cm} R^{(3)}=C_{23}\frac{W_{\lambda_{3},l+\frac{1}{2}}(\xi_{3})}{\xi_{3}}\ \ \ \hspace{1cm}r>b
  \end{eqnarray}
 The boundary conditions of the radial wave functions and their densities of current continuity at the interfaces $r=a$ and  $r=b$ require
\begin{eqnarray}
R^{(1)}(r)\frac{dR^{(2)}(r)}{dr}\mid_{r=a}&=&R^{(2)}(r)\frac{dR^{(1)}(r)}{dr}\mid_{r=a}\\
R^{(2)}(r)\frac{dR^{(3)}(r)}{dr}\mid_{r=b}&=&R^{(3)}(r)\frac{dR^{(2)}(r)}{dr}\mid_{r=b}
\end{eqnarray}
 By using the above two equations, one can obtain the eigenvalue $E$ of electron in MLQAD with donor
 impurity in the center.

 The oscillator strength is a very important physical quantity in the study of optical properties which related to the electronic transition. Using the obtained electron energies and normalized wave functions the oscillator strength of the intersubband transition between the states $\psi_{nlm}$ and
$\psi_{n'l'm'}$ in the dipole approximation, with the radiation of linearly polarized light along $z$-axis, are
given by the following expression \cite{davat}
\begin{eqnarray}
&&f_{nlm-n'l'm'} = \frac{2\mu}{\hbar^{2}}\Delta E_{fi} |\langle f|z|i\rangle|^{2}=\frac{2\mu}{\hbar^{2}}\Delta E_{fi} \times \nonumber \\
&&\left|\int R_{nl}R_{n'l'}r^3 dr\ \int {Y_{lm}Y^{\ast}_{l'm'}\cos\theta d\Omega}\right|^{2}.
 \end{eqnarray}
where, $\Delta E_{fi}$ denote the energy difference between the final and the initial
state energy and $|\langle f|z|i\rangle |^{2}$ is the transition matrix element that in second  line it has been expressed in terms of normalized
wave functions and spherical harmonics. The selection rules for such
transitions are determined by integrating over angular variables. With the substitution
$\cos\theta=(\sqrt{4\pi}/{3})Y_{1,0}$, it can be readily shown that the
probability of transition is unequal to zero only when
$|l-l'|=1$ and $m=m'$, where $l$ is the orbital, and $m$ is the magnetic quantum number. By using of this information and the layered nature of MLQAD,  one can rewrite the oscillator strength as
\begin{eqnarray}
 &&f_{nlm-n'l'm'}=\frac{2\mu}{\hbar^{2}}\Delta E_{fi}\times \nonumber \\
 && \left|\left(\int^{a}_{0} R_{nl}^{(1)}R_{n'l'}^{(1)}r^{3}dr +\int^{b}_{a} R_{nl}^{(2)}R_{n'l'}^{(2)}r^{3}dr +\int^{\infty}_{b} R_{nl}^{(3)}R_{n'l'}^{(3)}r^{3} dr\right)\right|^{2}\times \nonumber \\
 &&\left|\int {Y_{lm}Y^{\ast}_{l'm'}\cos\theta d\Omega}\right|^{2}.
\end{eqnarray}

\section{Results and discussion}
\subsection{Energy spectrum}
\label{sec:1}
\begin{figure}
\includegraphics[height=2.5in,width=3.2in]{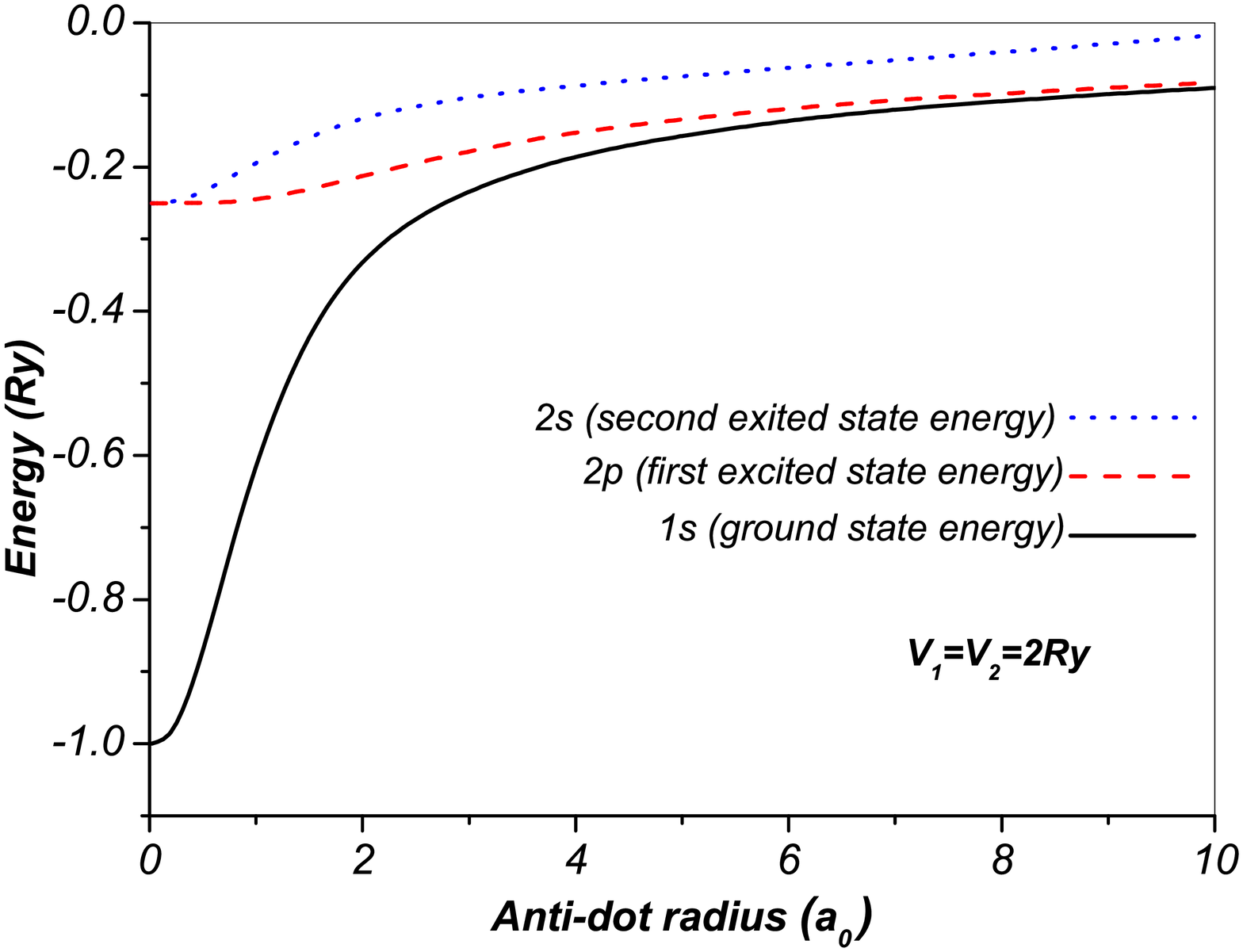}
\caption{The binding energies of the ground, first and  second excited state of the MLQAD hydrogenic impurity for the same confining core and shell potential ($V_{1}=V_{2}=2 Ry$) as a function of antidot radius for (a) finite CPs (b) infinite CPs.}
\label{f1}
\end{figure}

\begin{figure*}
\begin{center}
 \mbox{\includegraphics[height=2.5in,width=3.2in]{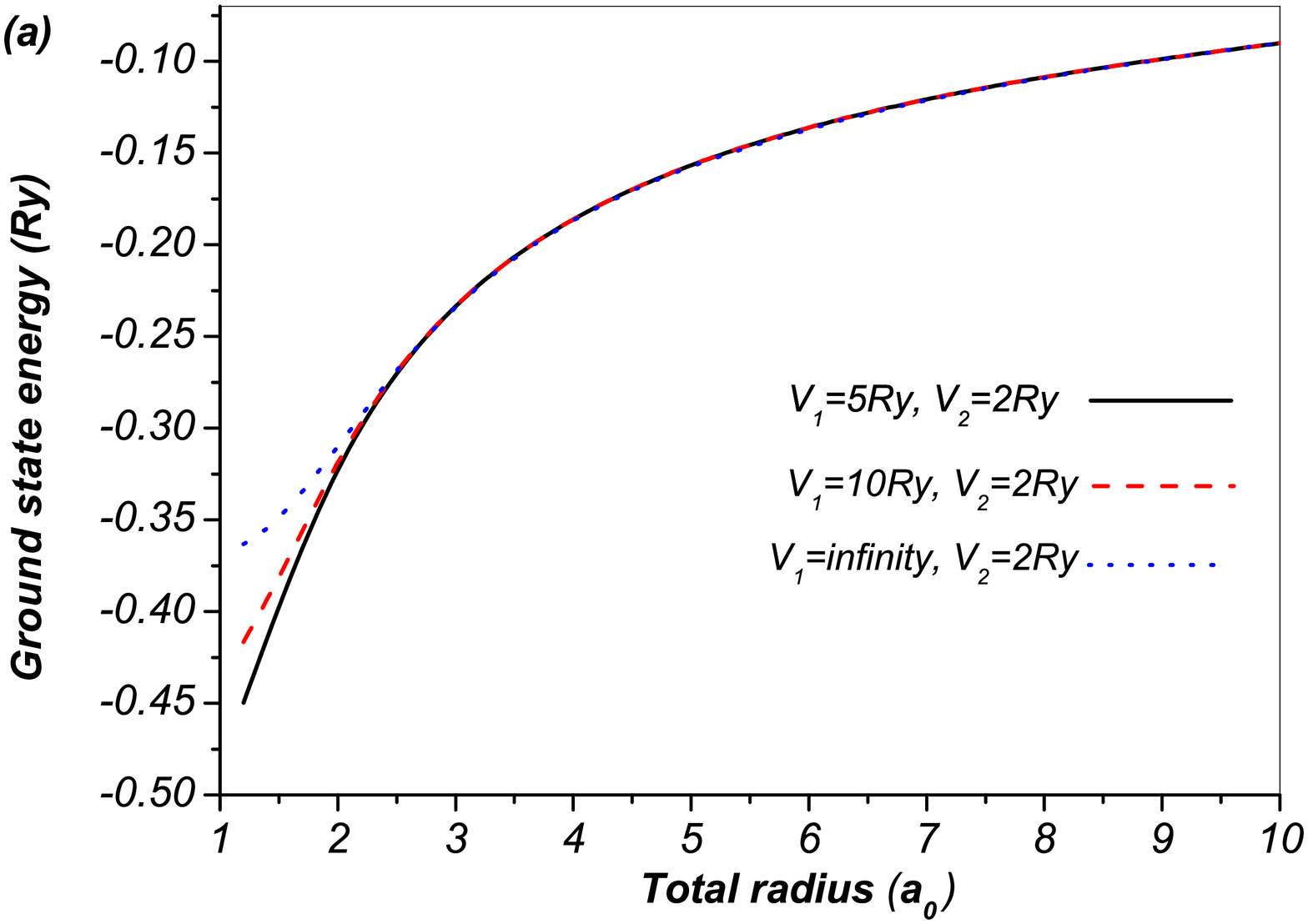}\quad
\includegraphics[height=2.5in,width=3.2in]{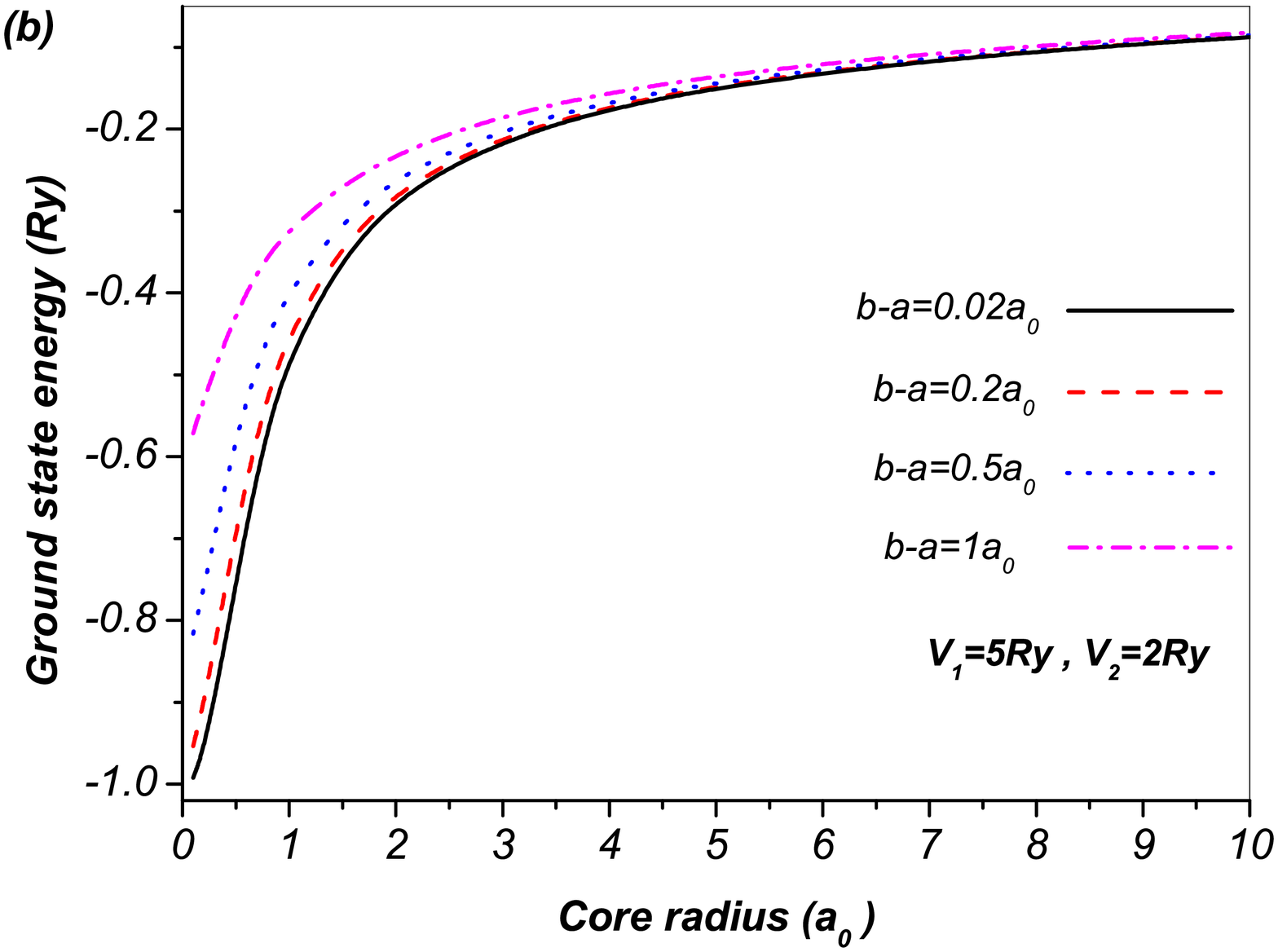} }
 \caption{The ground state energy (a) as a function of total antidot radius for various $V_{1}=5, 10, \infty Ry $ by choosing $a=a_{0}$ and $V_{2}=2 Ry$ (b) as a function of core radius for various shell thicknesses $b-a=0.02, 0.2, 0.5, 1 a_{0}$ with $V_{1}=5 Ry$ and $V_{2}=2 Ry$.}
\label{f2}\end{center}\end{figure*}

Here, the binding energies of the ground state and the first two excited states of a hydrogenic impurity
located at the centre of the MLQAD are calculated as a function of core radius for different shell thicknesses and CPs. To make a comparison, we focus on the special case of the same CPs (i.e., $V_{1}=V_{2}=V$). In fig.~\ref{f1}, the energy spectrum for $V_{1}=V_{2}=2Ry$ as a function of antidot radius are plotted. As we expected, the result are identical to the SLQAD that has been studied in \cite{davat,holo}. It is obvious from fig.~\ref{f1} that the orbital
energy degeneracy in $2s$ and $2p$ states, is removed. The energies for a free hydrogen atom is $-1/n^{2} Ry$, where $n$ is the principle quantum number, so the ground state ($1s$) and the first two excited state ($2s$ and $2p$) energies are equal to $-1 Ry$ and $-0.25 Ry$ respectively. From fig.~\ref{f1}, one can see that the energy curves will approach to the energy values of a free hydrogen atom in each state when the antidot radius becomes smaller and smaller. Although, when the radius of impurity increases, the confinement effect enhances the state energies and at very large radii, a bounded electron to an impurity never sees the surface of the antidot potential and behaves as a free electron.  \\
In the case of a MLQAD with different CPs ($V_{1}\neq V_{2}$), the ground state energy is plotted for two different cases in fig.~\ref{f2}. In fig.~\ref{f2} (a) the ground state energy vs total antidot radius is plotted when the shell CP and the core radius are fixed as $V_{2}=2Ry$ and $a=a_{0}$ whiles a core CP $V_{1}$ has three distinct values $5$, $10$ and $\infty Ry$. In fig.~\ref{f2} (b) the variation of the ground state energy with respect to the core radius is investigated for fixed values of CPs and shell thickness.
From fig.~\ref{f2} (a), it is clear that the binding energies are not influenced by the core CP value when the total radius is larger than about $2.5a_{0}$ but they are dependent to the core CP value when the total radius is less than $2.5a_{0}$. More precise, in the range of $a_{0}<b<2.5a_{0}$, a higher core CP leads to a larger binding energy.
fig.~\ref{f2} (b) shows that in the small core radii, the smaller shell thickness leads to the more bounded electron, while in large enough core radii (about $a>8a_{0}$) the results are independent on the shell thickness and the electron is no strongly bind in this case. Also, when the shell thickness becomes very small the ground state energy curve approaches to the SLQAD case (see fig.~\ref{f1} solid curve).

 \subsection {Radial probability distribution}
 \label{sec:2}
\begin{figure*}
\begin{center}
 \mbox{\includegraphics[height=2in,width=3in]{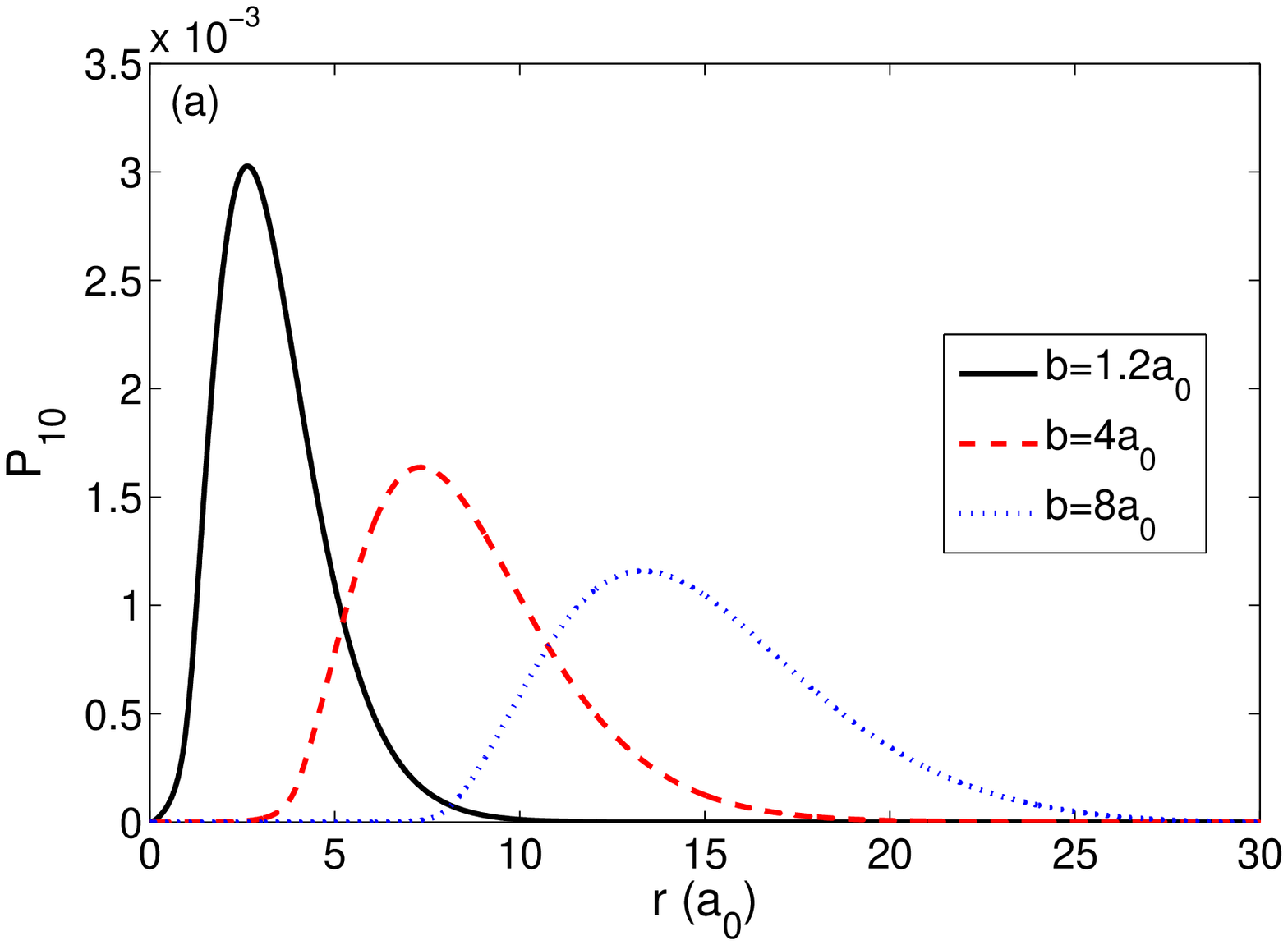}\quad
\includegraphics[height=2in,width=3in]{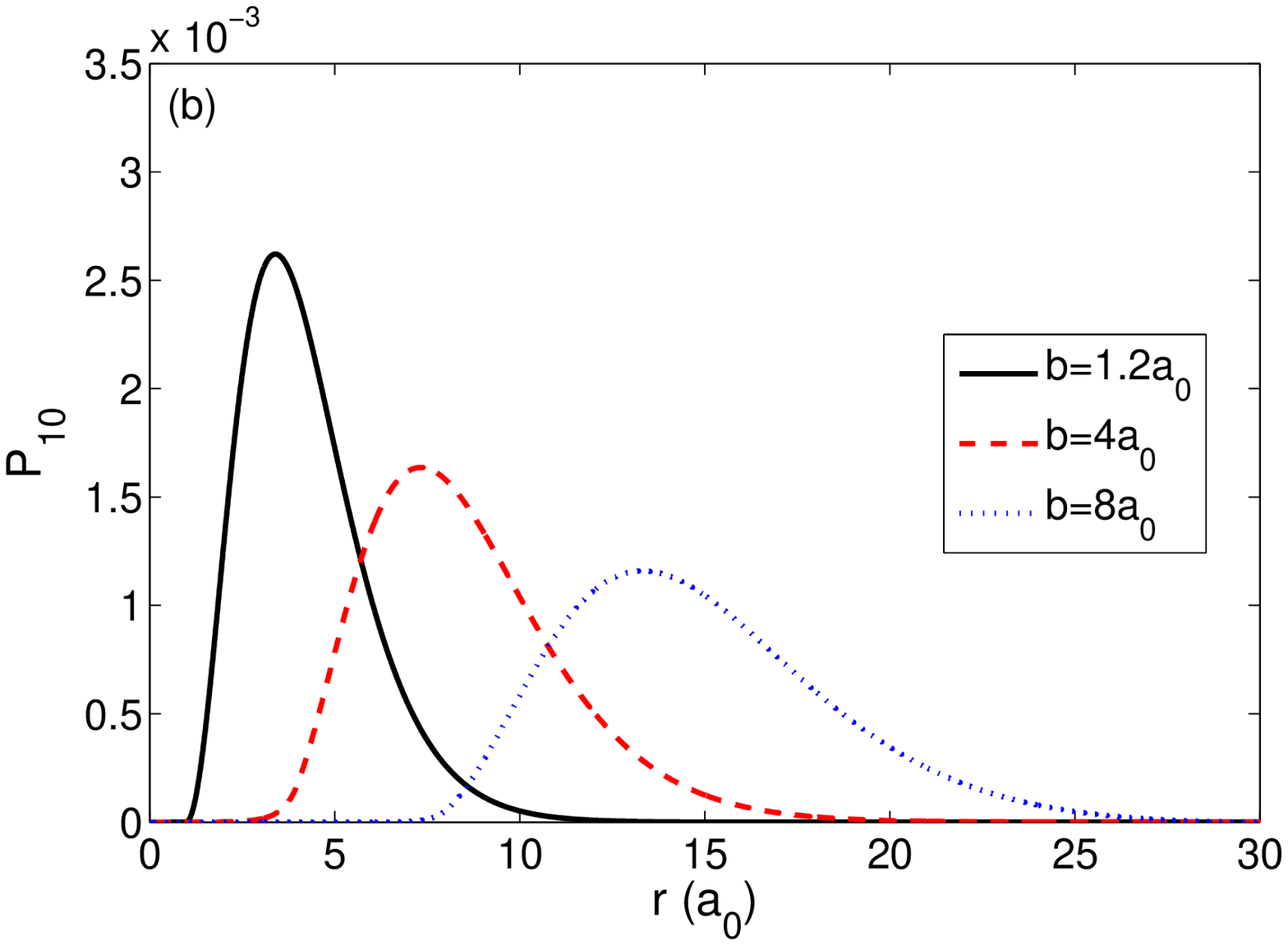} }
 \caption{ The radial probability distribution of the ground state ($P_{10}$) for several total radii with $a=a_{0}$ by choosing (a) $V_{1}=5Ry$ and $V_{2}=2Ry$ (b) $V_{1}=\infty$ and $V_{2}=2Ry$.}
\label{f3}\end{center}\end{figure*}

\begin{figure*}
\begin{center}
 \mbox{\includegraphics[height=2in,width=3in]{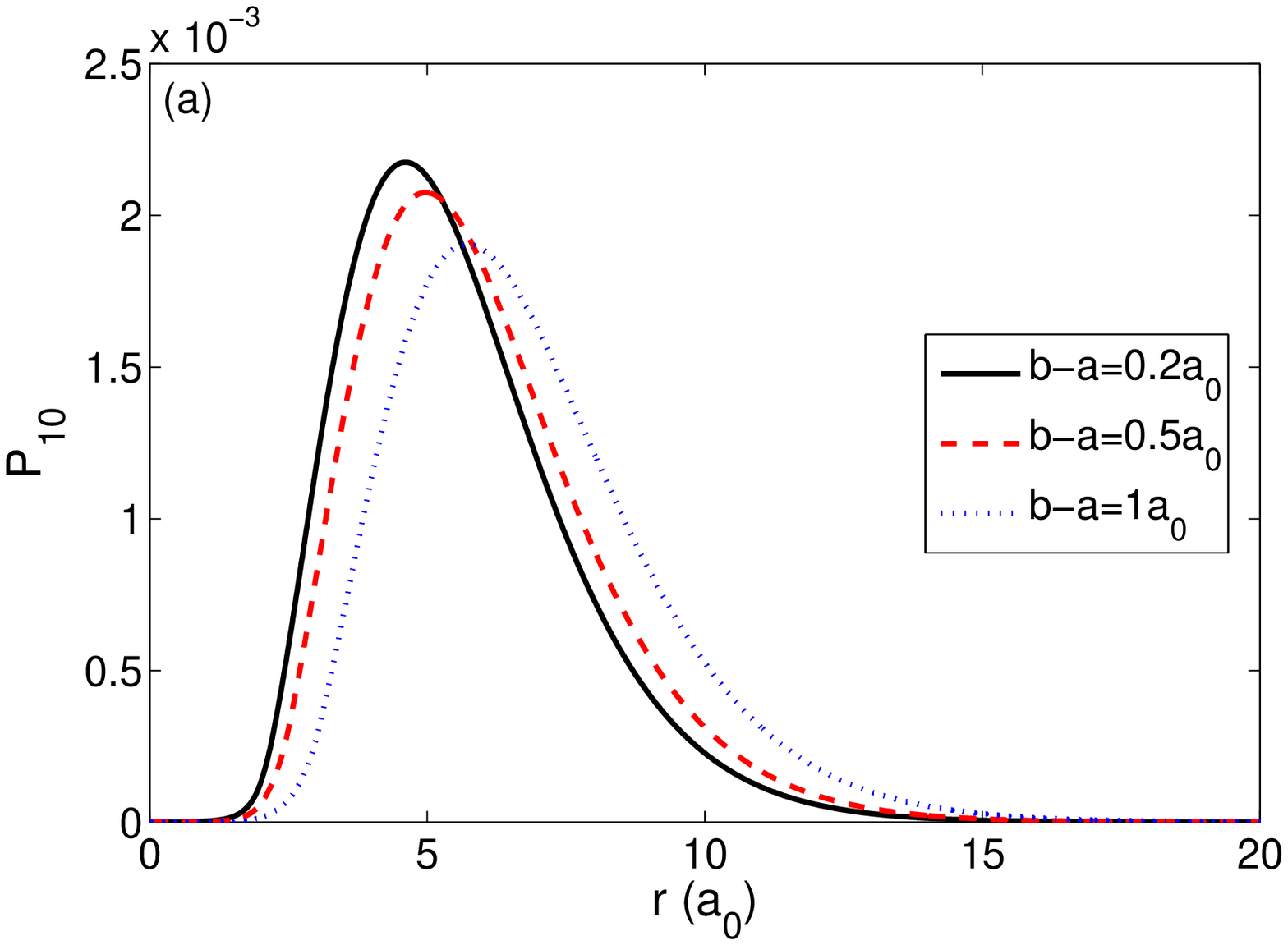}\quad
\includegraphics[height=2in,width=3in]{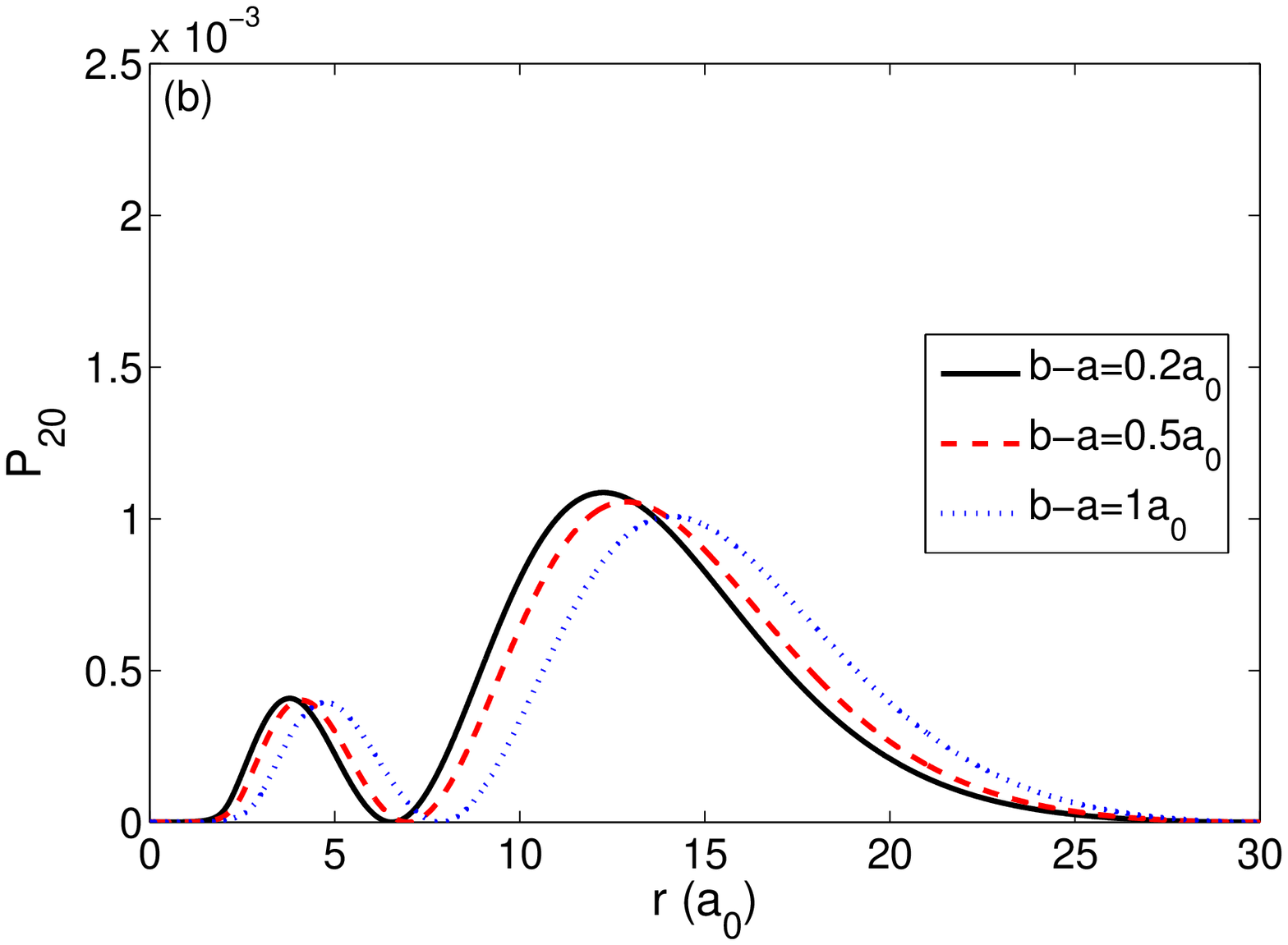} }
 \caption{The radial probability distribution  for various shell thicknesses $b-a=0.2, 0.5, 1 a_{0}$ with $a=a_{0}$, $V_{1}=5Ry$ and $V_{2}=2Ry$ for  (a) the ground state, $P_{10}$ (b) the second excited state, $P_{20}$.}
 \label{f4}\end{center}\end{figure*}

 In figs.~\ref{f3} (a) and (b) the radial probability distribution of the electron in its ground state, $P_{1s}$, for a couple of core and shell CPs and for three different total antidot radii, are plotted. The results for the radial probability distribution, clearly indicate that by increasing the total radius, the radial probability function becomes broader and expands to longer distances from the origin, in this sense, the electron becomes more delocalized. As the radial probability distribution gets broader, its height decreases, in accordance with the conservation of probability. By comparing the mutual curves (solid by solid and so on) we found that the core CP values have influenced mainly on the radial probability distribution of antidot with the small total radius (i.e., $b=1.2a_{0}$ or solid curves) and two other curves (dashed and dotted curve) are closely similar for both core CPs $V_{1}=5 Ry$ and $V_{1}=\infty$. The radial probability distribution of the ground and second excited state for various shell thicknesses are computed in figs.~\ref{f4} (a) and (b) respectively. These graphes indicate that by increasing the shell thickness the radial probability density function expanded to farther places from the origin. In the limiting case $(b-a)\rightarrow0$ the result approaches to the SLQAD case as expected.

\vspace{0.3cm}

\begin{figure*}
\begin{center}
 \mbox{\includegraphics[height=2.5in,width=3.2in]{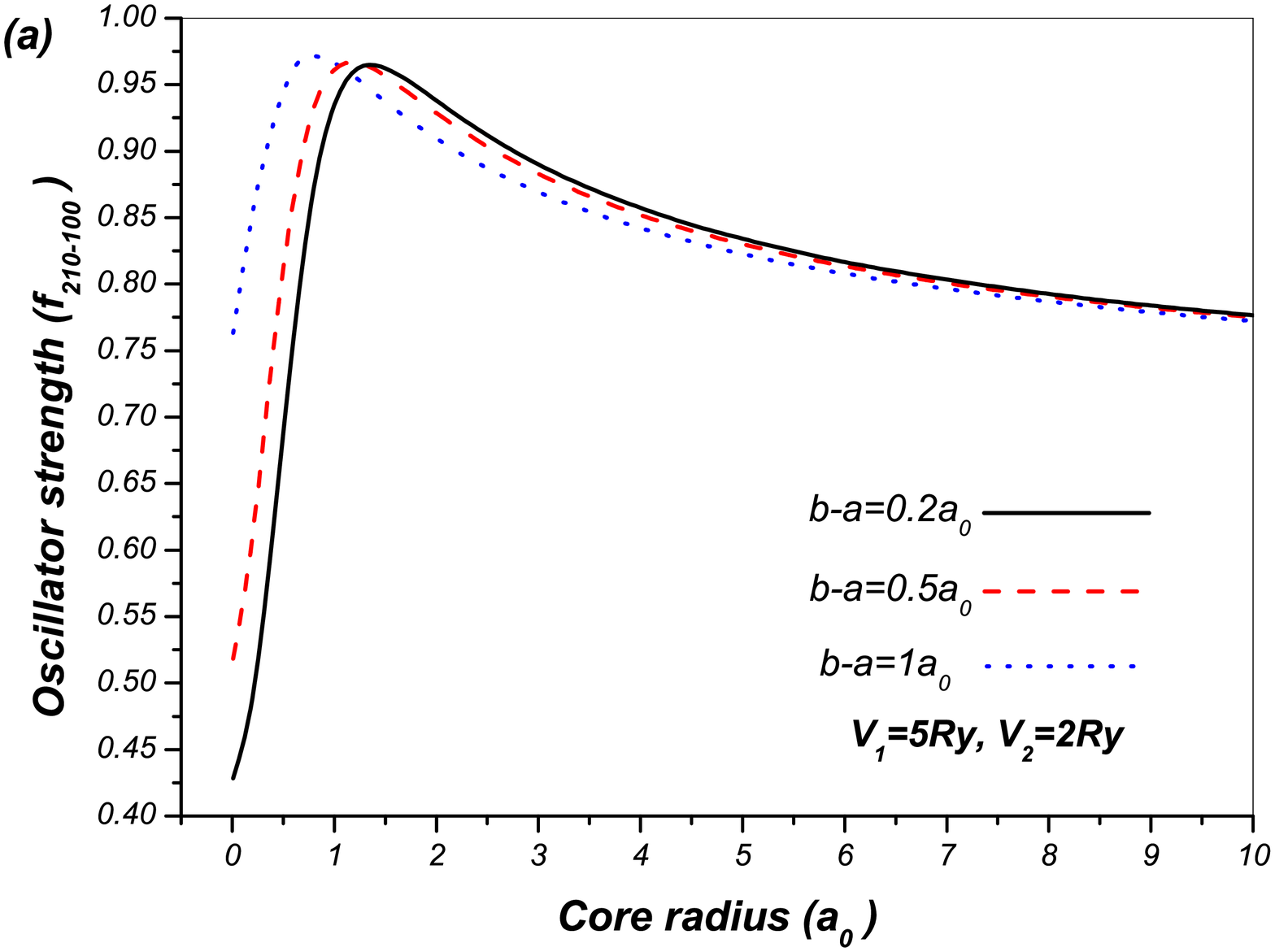}\quad
\includegraphics[height=2.5in,width=3.2in]{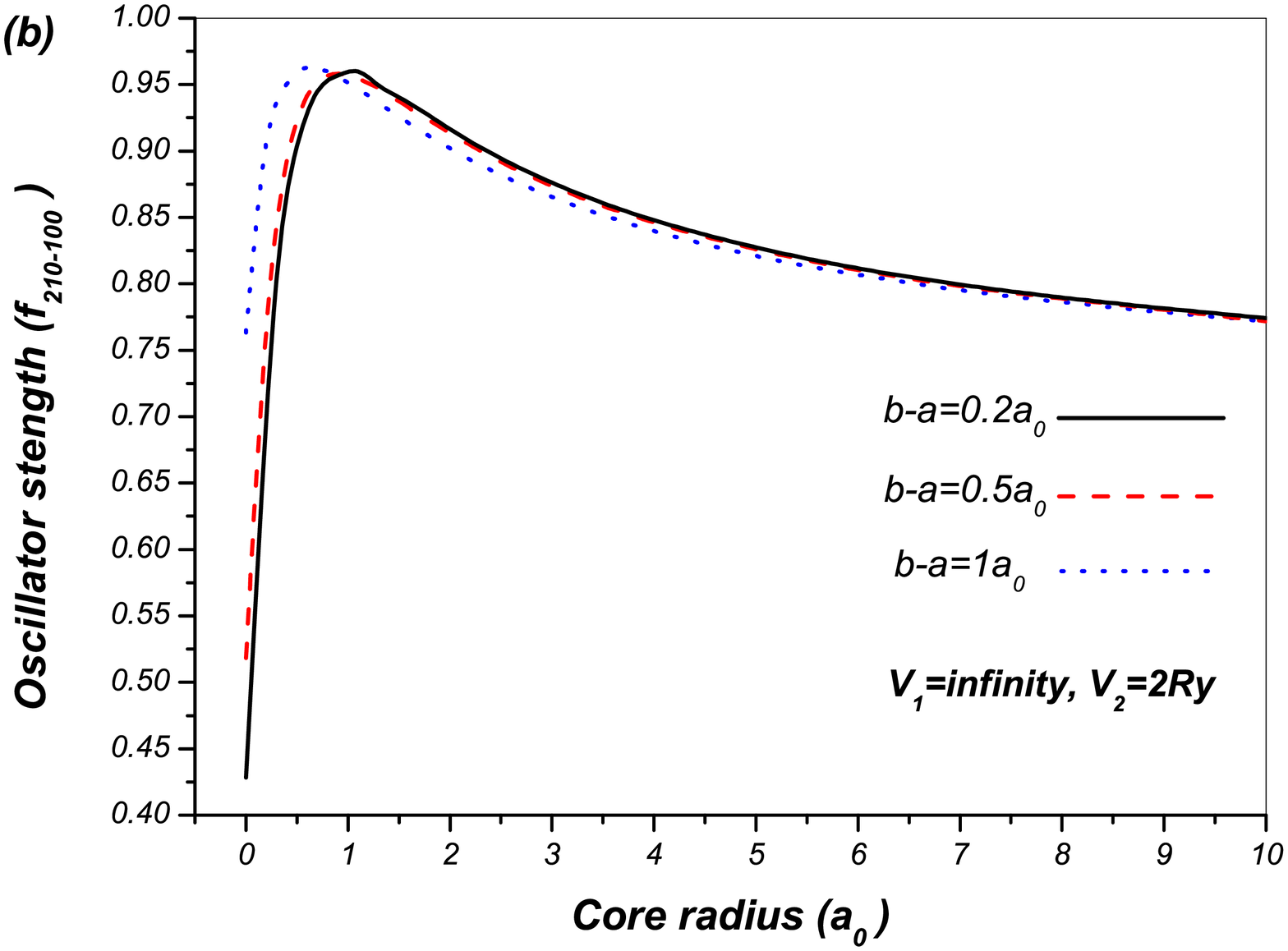} }
 \caption{Oscillator strength from the ground to the first allowed excited state ($f_{210-100}$) as a function of core radius for several shell thicknesses $b-a=0.2, 0.5, 1 a_{0}$ with (a) $V_{1}=5Ry$ and $V_{2}=2Ry$  (b) $V_{1}=\infty$ and $V_{2}=2Ry$.}
 \label{f5}\end{center}\end{figure*}
  \begin{figure*}
\begin{center}
 \mbox{\includegraphics[height=2.5in,width=3.2in]{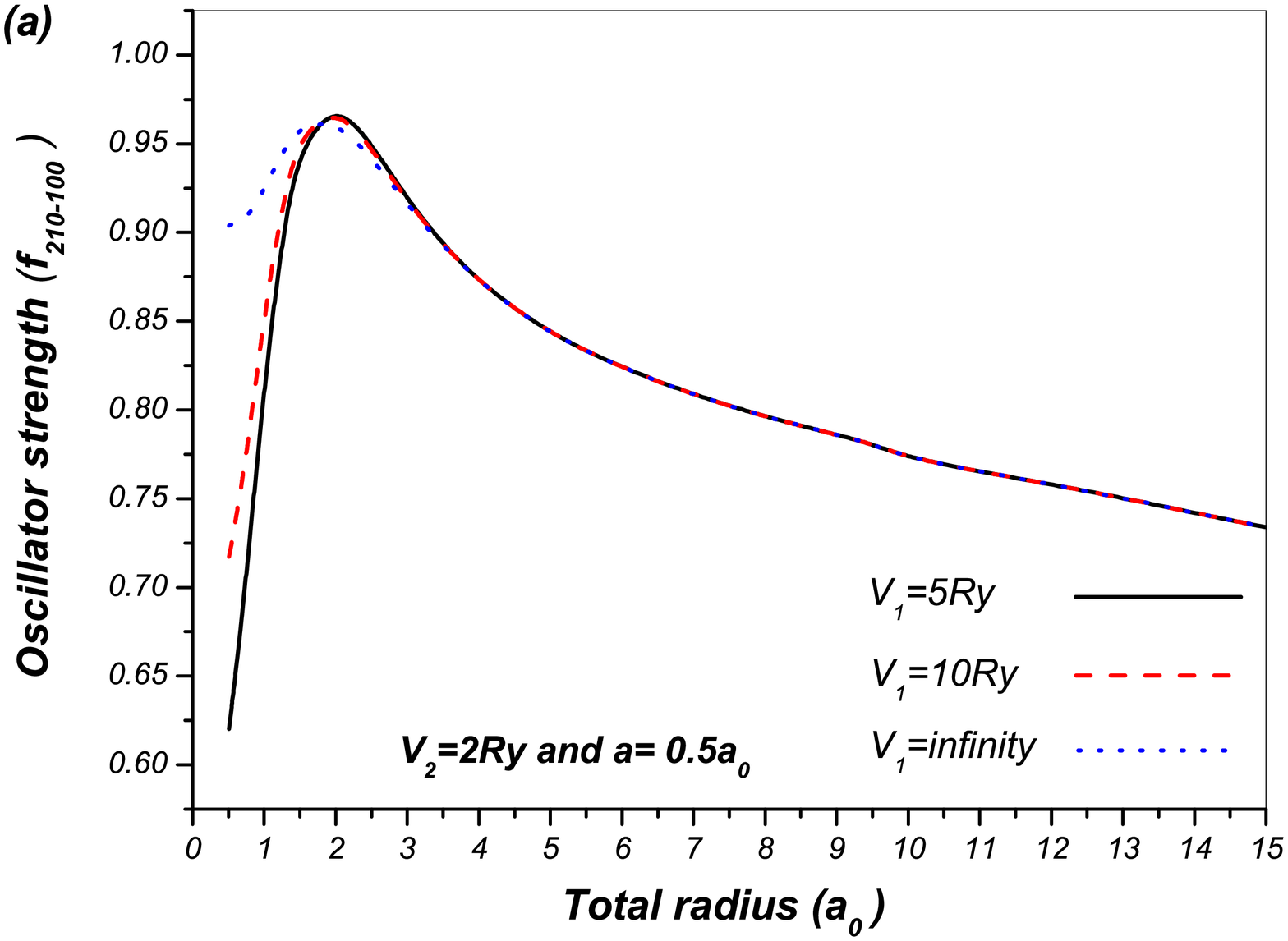}\quad
\includegraphics[height=2.5in,width=3.2in]{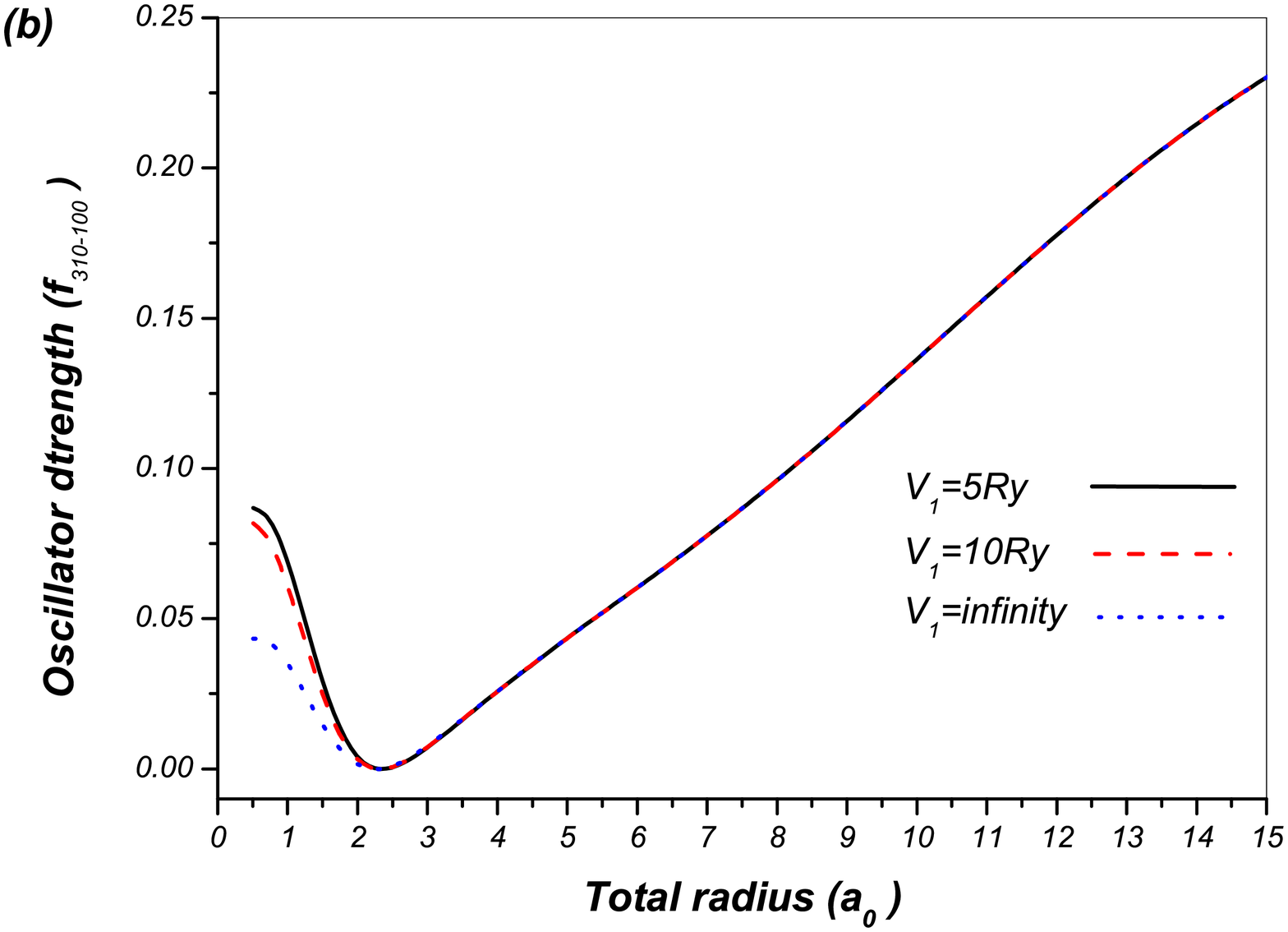} }
 \caption{Oscillator strength from the ground to  (a) the first allowed excited state ($f_{210-100}$) (b) the second allowed excited state ($f_{310-100}$) for several confining core potentials $V_{1}=5, 10, \infty Ry$ with constant shell potential $V_{2}=2Ry$ and $a=0.5a_{0}$ as a function of the total antidot radius.}
 \label{f6}\end{center}\end{figure*}

 \subsection{ Oscillator strength}
 \label{sec:3}
 The oscillator strength of an electron bound to a donor impurity for transition from the ground state to the first excited state, and for various shell thicknesses with finite and infinite core CP, are plotted as a function of core radius in figs.~\ref{f5} (a) and (b). As seen from figs, the oscillator strength curves undergo a peak in given core radii due to the big energy difference and good overlapping of the corresponding wave functions. When the value of shell thickness increases, the place of peak shifts toward smaller core radius and when the shell thickness becomes small ($b-a\rightarrow0$), the result (solid curve) approaches to the SLQAD case. By comparing the figs.~\ref{f5} (a) and (b) one can find that by increasing the core potential (from $5Ry$ to $\infty$) the difference among different shell thicknesses becomes smaller.

 In fig.~\ref{f6}, the variation of oscillator strength with respect to the total antidot radius for several core CPs have given  when the core radius and the shell CP are fixed as $a=0.5 a_{0}$ and $V_{2}=2 Ry$ respectively. According to the normalization condition, the summation of all allowed transition probabilities is $1$. From fig.~\ref{f6} (a), one can easily see that at the given core radius (about $a=2 a_{0}$), $f_{210-100}$ displays a maximum and obtains the major portion of $1$ (larger than $0.95$), so in this given radius the other transition probabilities have a very small portion of $1$ and tend toward $0$. Fig.~\ref{f6} (b) shows that at about $a=2 a_{0}$ that $f_{210-100}$ displays a maximum, $f_{310-100}$ displays a minimum and tends toward $0$. However, for large enough total antidot radii, the two transition probabilities tend to become comparable.
  
\section{conclusion}
The exact solution of Schr\"{o}dinger equation within the effective mass approximation are obtain for the electron in the MLQAD. It is found that only in the small total antidot radius a higher core CP leads to a larger binding energy. Also, it is shown that in the small core radii, the smaller shell thickness leads to the more bounded electron, while in large enough core radii (about $a > 8a_{0}$) the results are independent on the shell thickness. The oscillator strength for the intersubband electric transition from the ground state to the two first allowed excited state calculated by using the obtained spectrum and the corresponding wave functions. It found that for a fixed value of the core radius and shell potential by varying the core CP value, $f_{210-100}$ displays a maximum larger than $0.95$ at the given core radius (about $a=2 a_{0}$) so the transitions involving the second allowed state, $f_{310-100}$, can be neglected. But, in particular, for large enough total antidot radii, $f_{210-100}$ and $f_{310-100}$ become comparable. This indicates that the transitions involving the second allowed state, are no longer negligible when investigating the optical properties of the hydrogenic MLQAD with large enough total antidot radii, of the order of several hundred nanometers.


\vspace{0.3cm}


\begin{thebibliography}{99}
\bibitem{boga}  Bogachek, E.N., Landman, U.: Phys. Rev. B \textbf{52}, 14067 (1995)
\bibitem{aqui} Aquino, N., Castano, E., Koo, E.L.: Cin. J. Phys. \textbf{41}, 276 (2003)
\bibitem{khor} Khordad, R.: Solid State Sci. \textbf{12}, 1253 (2010)
\bibitem{plan} Planelles, J., Climente, J.I., Rajadell, F.: Physica E \textbf{33}, 370 (2006)
\bibitem{madhav} Madhav, A.V., Chakraborty, T.: Phys. Rev. B \textbf{49}, 8163 (1994)
\bibitem{holo} Holovatsky, V.A., Makhanets, O.M., Voitsekhivska, O. M.:  Physica E \textbf{41}, 1522 (2009)
\bibitem{davat} Davatolhagh, S., Jafari, A.R., Vahdani, M.R.K: Superlattices and Microstruchure \textbf{51}, 62 (2012)
\bibitem{kohn} Kohn, W.: Solid State Phys. \textbf{5}, 257 (1957)
\bibitem{hung} Haug H., Koch, S.W.: Quantum theory of the Optical and Electronic Properties of SemiConductors, third ed., World Scienti.c, Singapore, (1994)
\bibitem{bas1} Bastard, G.: Surf. Sci. \textbf{113}, 165 (1982)
\bibitem{bas2} Bastard, G.: Phys. Rev. B \textbf{24}, 4714 (1983)
\bibitem{safak} Yilmaz, S., Safak, H.: Physica E  \textbf{36}, 40 (2007)
\bibitem{ozmen} Ozmen, A., Yakar, Y., Cakir, B., Atav, U.: Optic Communications \textbf{282}, 3999 (2009)
\bibitem{sadeghi} Sadeghi, E.: Physica E \textbf{41}, 1319 (2009)
\bibitem{osorio} Oosorio, F.A.P., Marques, A.B.A, Machado, P.C.M, Borges, A.N: Microelectronics journal \textbf{36}, 244 (2005)
\bibitem{siva} Sivakami, A., Mahendran, M.: Physica B  \textbf{405}, 1403 (2010)
\bibitem{varshini} Varshni, Y.P.: Physics Letters A \textbf{252}, 248 (1999)
\bibitem{poras} Porras-Montenegro, N., Perez-Merchancano, S.T.: Phys. Rev. B  \textbf{46}, 9780 (1992)
\bibitem{Chuu} Chuu, C. M., Hsiao, W.N., Mei, W.N.: Phys. Rev \textbf{46}, 3898 (1992)
\bibitem{tkac1} Tkach, M., Holovatsky, V., Berezovsky, Ya.: Physics and Chemistry Solid State (ukr) \textbf{4}, 213 (2003)
\bibitem{tkac2} Tkach, M., Holovatsky, V., Berezovsky, Ya.: Bulletin of the Russian Academy of Sciences: Physics. \textbf{68}, 120 (2004)
\bibitem{zu} Zu, J.L., Xiong, J.J., Gu, B.L.: Phys. Rev. \textbf{41}, 6001 (1990)
\bibitem{cheng} Hsieh, C.Y., Chuu, D.S.: J. Phys. Condens. Matter \textbf{12}, 5641 (2000)
\bibitem{ada} Adachi, S.: J. Appl. Phys. \textbf{58}, R1 (1985)
\bibitem{zet} Zettili, N.: Quantum mechanic, consept and application, John Wiley and Sons Ltd,  pp. 287-289 (2001)
\bibitem{abra} Abramowitz, A., Stegun,I.: Handbook of Mathematical Function with Formulas, Graphs and Mathematical Tables, US GPO,
Washington, DC, pp. 505-509 (1994)
\end{thebibliography}
\end{document}